# A New Heavy-Fermion Superconductor CeIrIn$_5$: Relative of the Cuprates?

C. Petrovic[1,2], R. Movshovich[1], M. Jaime[1], P. G. Pagliuso[1], M. F. Hundley[1], J. L. Sarrao[1], Z. Fisk[1,2], and J. D. Thompson[1](*)

[1] Condensed Matter and Thermal Physics, Los Alamos National Laboratory, Los Alamos, NM 87545 USA.

[2] National High Magnetic Field Laboratory, Florida State University, Tallahassee, FL 32306 USA

* To whom correspondence should be addressed. E-mail: jdt@lanl.gov

**Abstract.** - CeIrIn$_5$ is a member of a new family of heavy-fermion compounds and has a Sommerfeld specific heat coefficient of 720 mJ/mol-K$^2$. It exhibits a bulk, thermodynamic transition to a superconducting state at $T_c$=0.40 K, below which the specific heat decreases as $T^2$ to a small residual T-linear value. Surprisingly, the electrical resistivity drops below instrumental resolution at a much higher temperature $T_0$=1.2 K. These behaviors are highly reproducible and field-dependent studies indicate that $T_0$ and $T_c$ arise from the same underlying electronic structure. The layered crystal structure of CeIrIn$_5$ suggests a possible analogy to the cuprates in which spin/charge pair correlations develop well above $T_c$.

Of the vast number of metallic compounds, only a small fraction enter a superconducting state at low temperatures, and of this small number, an even smaller fraction develop superconductivity out of a normal state in which electronic correlations produce orders-of-magnitude enhancement of the conduction electrons' effective mass [1]. This subset of materials, known as heavy-fermion superconductors, has been an influential area of research in condensed matter physics since its first member $CeCu_2Si_2$ was discovered [2] in 1979. Unlike all previously known superconductors, the presence of a magnetic ion (in this case Ce) was essential for superconductivity and the temperature dependence of physical properties below the superconducting transition temperature $T_c$ was inconsistent with the well-established Bardeen-Cooper-Schrieffer theory of superconductivity. Over the past two decades other examples have been added to this class: five uranium-based compounds at atmospheric pressure and five cerium-based systems in which heavy-fermion superconductivity has been induced by applying pressure [3]. Interestingly, all but two of the pressure-induced heavy-fermion superconductors and one U-based superconductor form in the same $ThCr_2Si_2$ tetragonal structure as $CeCu_2Si_2$, suggesting that this structure type is particularly favorable for heavy-fermion superconductivity. The two notable Ce-based exceptions have a common denominator as well, $CeIn_3$. Cubic $CeIn_3$, when subjected to 25-kbar pressure, becomes a superconductor below about 0.15K [4], and $CeRhIn_5$, composed of layers of $CeIn_3$, superconducts below 2.1K for pressures above 17 kbar [5]. These two recent discoveries suggest that heavy-fermion superconductivity might be found in structurally-related materials.

Experimental and theoretical study of the superconductivity in these heavy-fermion materials has formed a substantial basis for understanding more broadly classes of unconventional superconductors, including the high-$T_c$ cuprates, in which the electron-pairing interaction responsible for superconductivity may be mediated by spin fluctuations [1]. In spite of progress, the heavy-fermion problem and heavy-fermion superconductivity in particular remain challenges to experiment and theory [6]. Though heavy-fermion behavior has been found in several structure types, it appears that, like conventional BCS superconductivity, heavy-fermion superconductivity may be favored by particular crystallographic structures. Because of the limited number of examples, we know very little about relationships that should exist between the structure and properties of these materials. Any predictive understanding of how superconductivity can emerge in the highly correlated ground state has to be able to explain why it appears in one crystal structure and not another. This makes the discovery of a new prototype structure for heavy-fermion superconductivity of special interest. Here, we report a new ambient-pressure Ce-based heavy-fermion superconductor that is isostructural to $CeRhIn_5$, suggesting that this structure, like the $ThCr_2Si_2$ structure, may be particularly favorable for superconductivity. Unlike $CeCu_2Si_2$, this new compound grows easily and reproducibly as large, very pure single crystals, opening the possibility for unprecedented study.

$CeIrIn_5$ is a member of this new family that forms as $R_nT_mIn_{3n+2m}$, where n=1 or 2, m=1, R= La through Gd (except Eu), and T is a transition metal. All members grow readily as cm-sized, plate-like single crystals out of an In-rich flux. Crystals were obtained by combining stoichiometric amounts of Ce and T with excess In in an alumina crucible, encapsulating the crucible in an evacuated quartz ampoule, heating to 1100 C, and slowing cooling to 600 C. At this temperature, the ampoule was removed from the oven and the excess In removed by centrifugation. Powder x-ray patterns obtained on crushed single crystals show that $CeIrIn_5$ crystallizes in the tetragonal $HoCoGa_5$ structure type, with a=4.668(1) Å and c=7.515(2) Å [7]. Within typical resolutions of x-ray diffraction, microprobe analysis, and differential scanning calorimetry, the $CeIrIn_5$ crystals are single-phase. Compounds for which n=1 can be viewed as alternating layers of $CeIn_3$ and $TIn_2$ stacked sequentially along the tetragonal c-axis and for n=2 form as bilayers of $CeIn_3$ separated by a single layer of $TIn_2$. A large resistivity ratio $\rho(300 K)/\rho(2 K)$= 50-80 for the Ce-materials attests, in part, to the high quality of the crystals as does the observation of resolution-limited Laue diffraction and NQR spectra. The hallmark of a heavy-fermion system is the magnitude of its electronic coefficient of specific heat $\gamma$, which is a measure of the effective mass enhancement of conduction electrons produced by electronic correlations [1]. All of the Ce-based members of this new family exhibit heavy-fermion behavior as indicated by their large Sommerfeld specific heat coefficients $\gamma$, which range from $\approx$400 mJ/mole-Ce $K^2$ for antiferromagnetic $CeRhIn_5$ and $Ce_2RhIn_8$ to $\approx$700 mJ/mole-Ce $K^2$ for $CeIrIn_5$ and $Ce_2IrIn_8$. In contrast, the La-analogues, which do not contain an f-electron, are Pauli paramagnets with coefficients $\gamma$ of about 5 mJ/mole-$K^2$ that are typical of simple metals. Additional details of the preparation and characterization of this family will be given elsewhere.

The overall temperature dependence of the resistivity $\rho$ and magnetic susceptibility $\chi$ of $CeIrIn_5$ is shown in Fig. 1. The magnetic susceptibility is anisotropic, with $\chi$ larger by nearly a factor of two at low temperatures for a magnetic field applied along the tetragonal c-axis. Plots of $1/\chi$ are linear in temperature for T≥ 200K. From the linear regime, we find a paramagnetic Curie temperature $\Theta_P$, which is +12.5 K (–67.4 K) for a magnetic field of 1 kOe applied parallel (perpendicular) to the c-axis. A polycrystalline average of the high temperature data gives an effective moment $\mu_{eff}$ =2.28 $\mu_B$ that is reduced somewhat from the free-ion moment of $Ce^{3+}$, 2.54 $\mu_B$, due to the presence of crystalline electric fields that lift the degeneracy of the J=5/2 Hund's-rules multiplet. Characteristic of Ce-based heavy-fermion compounds, the resistivity passes through a maximum at low temperatures that typically is attributed to the cross-over from strong, incoherent scattering of electrons at high temperatures to the development of strongly correlated bands at low temperatures. The magnitude and temperature dependence of $\rho$ are similar to those of the heavy-fermion antiferromagnet $CeIn_3$ [4].

Thermodynamic and transport properties of $CeIrIn_5$ at low temperatures are summarized in Fig. 2. Above 0.4 K, the specific heat divided by temperature C/T≡ $\gamma$=720 mJ/mole-$K^2$ and is nearly temperature independent. At $T_c$=0.40 K, there is a jump in C/T and a prominent signature in ac susceptibility $\chi_{ac}$. Comparing the magnitude of this $\chi_{ac}$ response to that of a piece of superconducting tin having a similar size and shape as the $CeIrIn_5$ sample, we estimate that the $\chi_{ac}$ signature corresponds to a change in susceptibility of $-(1\pm 0.1)/4\pi$, as expected for a bulk superconductor. From the average of measurements on three different crystals, the specific heat jump $\Delta C$ at $T_c$ is equal to $(0.76 \pm 0.05)\gamma T_c$. This ratio $\Delta C(T_c)/\gamma T_c$ is comparable to that found in other heavy-fermion superconductors, such as $CeCu_2Si_2$ and $UPt_3$ [6], and provides compelling evidence that superconductivity in $CeIrIn_5$ develops among the heavy quasiparticles. The specific heat data below $T_c$ fit well to the sums of nuclear-Schottky, $T^2$ and T-linear contributions. (A nuclear Schottky term is expected due to the large nuclear quadrupole moments of Ir and In. [8]) The $C \propto T^2$ contribution suggests that the superconducting gap function goes to zero along certain portions of the Fermi surface [9]. The temperature dependence of the thermal conductivity, which is insensitive to the nuclear Schottky, also is described well from $T_c$ to 50 mK by the sum of linear and quadratic terms that are consistent with corresponding terms in the specific heat.

A peculiar aspect of the data in Fig. 2 is that the resistivity drops to zero, or at least to less than our instrumental resolution of 0.01μΩ-cm, at $T_0$=1.2 K without a prominent thermodynamic or magnetic signature [10]. As shown in Fig. 3, measurements of the specific heat, ac susceptibility and electrical resistivity in magnetic fields applied parallel and perpendicular to the c-axis of $CeIrIn_5$ find that the anisotropic responses of $T_c$, determined by specific heat and ac susceptibility, and $T_0$, determined resistively, are identical. Within the scatter of data in Fig. 3, these results are reproduced in three independently-grown crystals. It is extremely improbable that a secondary phase imbedded in each of these crystals would exhibit precisely the same anisotropy as the bulk phase below $T_c$, and, therefore, seems reasonable to conclude that both transitions at $T_0$ and $T_c$ are intrinsic and arise from a common underlying electronic structure that band structure calculations [11] and preliminary de Haas-van Alphen measurements [12] show to be quasi-2D. Though coming from a common electronic background, $T_c$ and $T_0$ develop out of apparently dissimilar manifestations of the highly correlated normal state. Just above $T_c$, the large, nearly constant C/T is typical of a strongly correlated Landau Fermi liquid. However, in all crystals we have studied, the electrical resistivity varies as $\rho(T)-\rho(T_0) \propto T^n$, with n=1.3± 0.05, for $T_0 \leq T \leq 5$ K. This is not the quadratic temperature dependence expected of a Landau Fermi liquid, and it persists unchanged from ~5 K to 60 mK when a magnetic field is applied to suppress $T_0$. Similar power-law variations in the electrical resistivity are found in the cuprates [13] and in heavy-fermion systems tuned by pressure to a magnetic/superconducting boundary [14]. One suggestion for its origin is the scattering of conduction electrons by antiferromagnetic spin fluctuations whose characteristic wave-vector connects portions of the Fermi surface [15]. In this scenario, the lack of a detectable change in the power-law indicates that a modest field does not substantially alter the nature of antiferromagnetic fluctuations or Fermi-surface topology of $CeIrIn_5$.

Tuning the hybridization between the 4f and ligand electrons by substituting Rh for Ir induces small-moment, incommensurate antiferromagnetism [16] in the end member $CeRhIn_5$, which has a Néel temperature of 3.8 K. Magnetization and nuclear quadrupole-resonance studies indicate similarities [5,16] of the magnetism in $CeRhIn_5$ with that found in $La_2CuO_4$ from which high-$T_c$ superconductivity develops with hole doping. When Rh is added substitutionally into $CeIrIn_5$, $T_0$ decreases and $T_c$ increases, which it also does when $CeIrIn_5$ is subjected to hydrostatic pressure. For x=0.25 and 0.5 in $CeIr_{1-x}Rh_xIn_5$, a small, less than 0.01(-1/4π), diamagnetic response appears in $\chi_{ac}$ at $T_0$ that is followed by a much larger diamagnetic signal at $T_c$. Midway between the end-points, $CeIr_{0.5}Rh_{0.5}In_5$, $T_c$ more than doubles to 0.86K, $\Delta C(T_c)/\gamma T_c$, $\gamma$, and $C(T)$ below $T_c$ are virtually unchanged relative to $CeIrIn_5$, and $T_0$ decreases to 1.0 K. Higher Rh concentrations (x≥0.6) induce a well-defined magnetic transition in both specific heat and magnetic susceptibility. These trends with isoelectronic substitution, which tunes f-d hybridization, are similar to those observed in the cuprates [17] with hole doping, which tunes band filling.

The apparently zero-resistivity state below $T_0$ suggests the presence of a percolating path of superconductivity along the sample. Though possible, it seems unlikely, given the high quality of the crystals and reproducibility of the effect, that the transition at $T_0$ arises from chemical or residual stress inhomogeneities in the sample. A possible alternative interpretation comes from an analogy with the cuprates. As a function of temperature and doping in the cuprates, a decrease in, for example, spin susceptibility and electrical resistivity defines a boundary in the T-x phase diagram that marks a cross-over from a paramagnetic to a pseudo-gap state out of which bulk superconductivity develops [18]. Many experiments are consistent with the formation of local (static or dynamic) spin/charge pair correlations without global phase coherence, concepts for which there is growing theoretical support [17]. In the cuprates, this boundary is smeared by inhomogeneity introduced by hole doping. Without such inhomogeneity, one might expect this boundary to become a sharply defined phase transition [19] with signatures similar to those found in our case at $T_0$. We, however, would expect a specific heat feature at $T_0$, but one is not prominent. Quite plausibly this feature is small compared to the large, heavy-electron specific heat out of which it develops and not readily detected within our experimental resolution. [10] Again drawing on the cuprates, the bulk transition at $T_c$ might be interpreted as the (Bose) condensation of electron pairs 'preformed' at $T_0$ or as the temperature at which Josephson coupling among pairs produces global phase coherence throughout the sample.

In summary, the new superconductor CeIrIn$_5$ suggests that the physics of heavy-fermion materials is much richer than previously imagined and that, when crystallizing in a quasi-2D structure, may show features analogous to those in the cuprate superconductors [20]. The search for yet other examples of 2-D structure types that form with Ce appears to be a fruitful path of investigation as does additional study of CeIrIn$_5$, which may bridge our understanding of more nearly 3-D heavy-fermion metals and the copper oxides.

\*\*

Work at Los Alamos was performed under the auspices of the U. S. Department of Energy. Z. F. acknowledges support through NSF grants DMR-9870034 and DMR-9971348.

Figure Legends:

Fig. 1 - (a). Magnetic susceptibility χ as a function of temperature for a 1-kOe field applied parallel to the c- (circles) and a-axis (triangles) of $CeIrIn_5$. Measurements were made in a Quantum Design $\underline{S}$uperconducting $\underline{Qu}$antum $\underline{I}$nterference $\underline{D}$evice magnetometer. (b). Electrical resistivity ρ versus temperature measured with a 4-lead ac resistance bridge.

Fig. 2 - Specific heat divided by temperature C/T (circles, left ordinate), ac magnetic susceptibility $χ_{ac}$ (triangles, arbitrary units) and electrical resistivity ρ (squares, right ordinate) of $CeIrIn_5$ as functions of temperature. The solid line is a fit to C/T data below half of $T_c$ and is given by C/T (J/mole $K^2$)= 3.9(±0.1)•$10^{-4}/T^3$ + 3.5(±0.1)T + 0.03(±0.2). The dashed line through $χ_{ac}$ (T) is a guide to the eye.

Fig. 3 - Magnetic field H versus temperature phase diagram constructed from specific heat C, ac magnetic susceptibility $χ_{ac}$ and electrical resistivity ρ measurements on a $CeIrIn_5$ crystal with a magnetic field applied parallel and perpendicular to the c-axis. Transition midpoints are used to define the diagram. The left ordinate and bottom abscissa correspond to resistivity data. The right ordinate and top abscissa correspond to specific heat and $χ_{ac}$ data. Open symbols are for H parallel to the c-axis. Solid symbols are for H perpendicular to the c-axis. Note the difference in field and temperature scales and that the anisotropy in these data are identical irrespective of the measurement technique. Solid and dashed lines are guides to the eye.

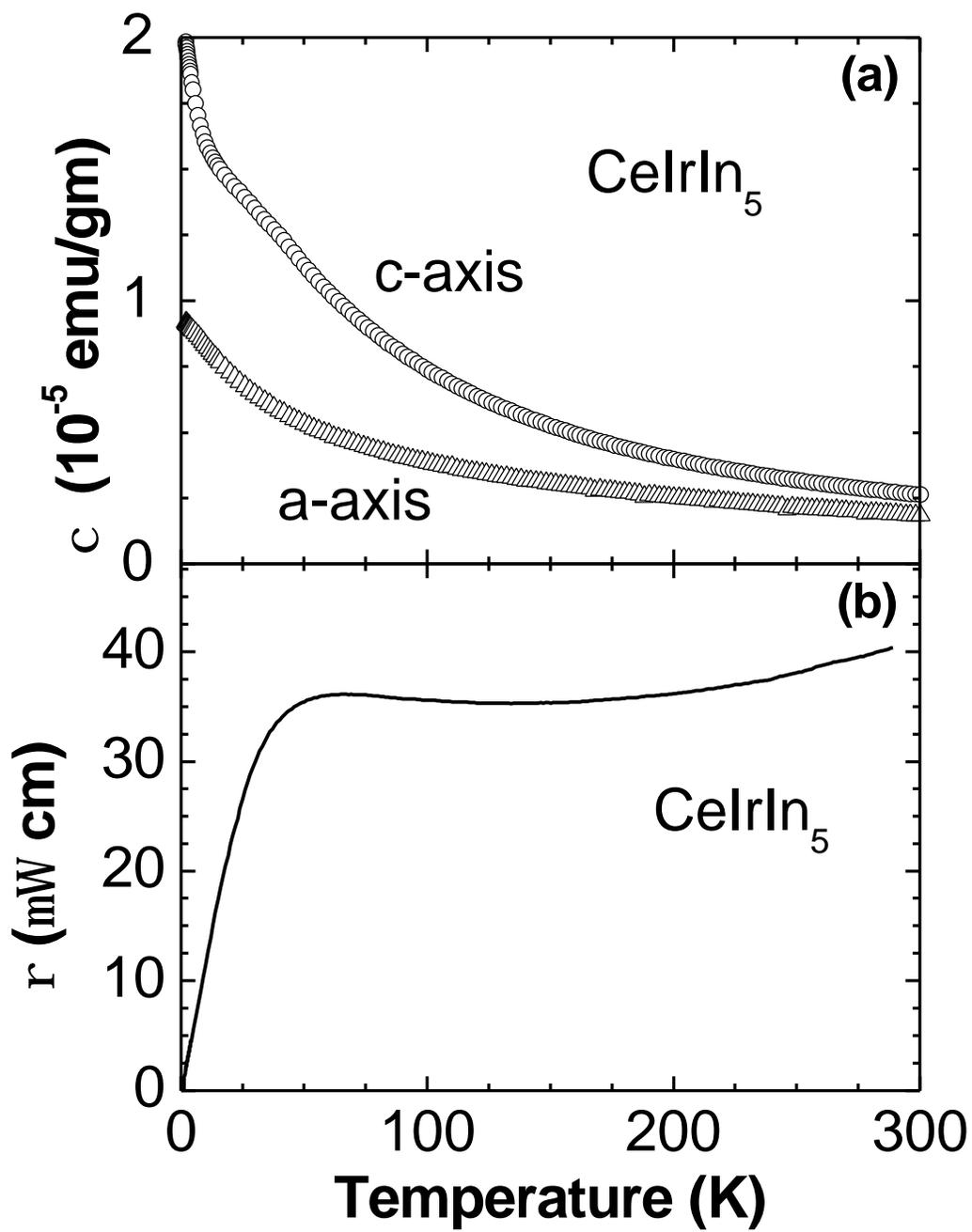

Figure 1 Petrovic, *et al.*

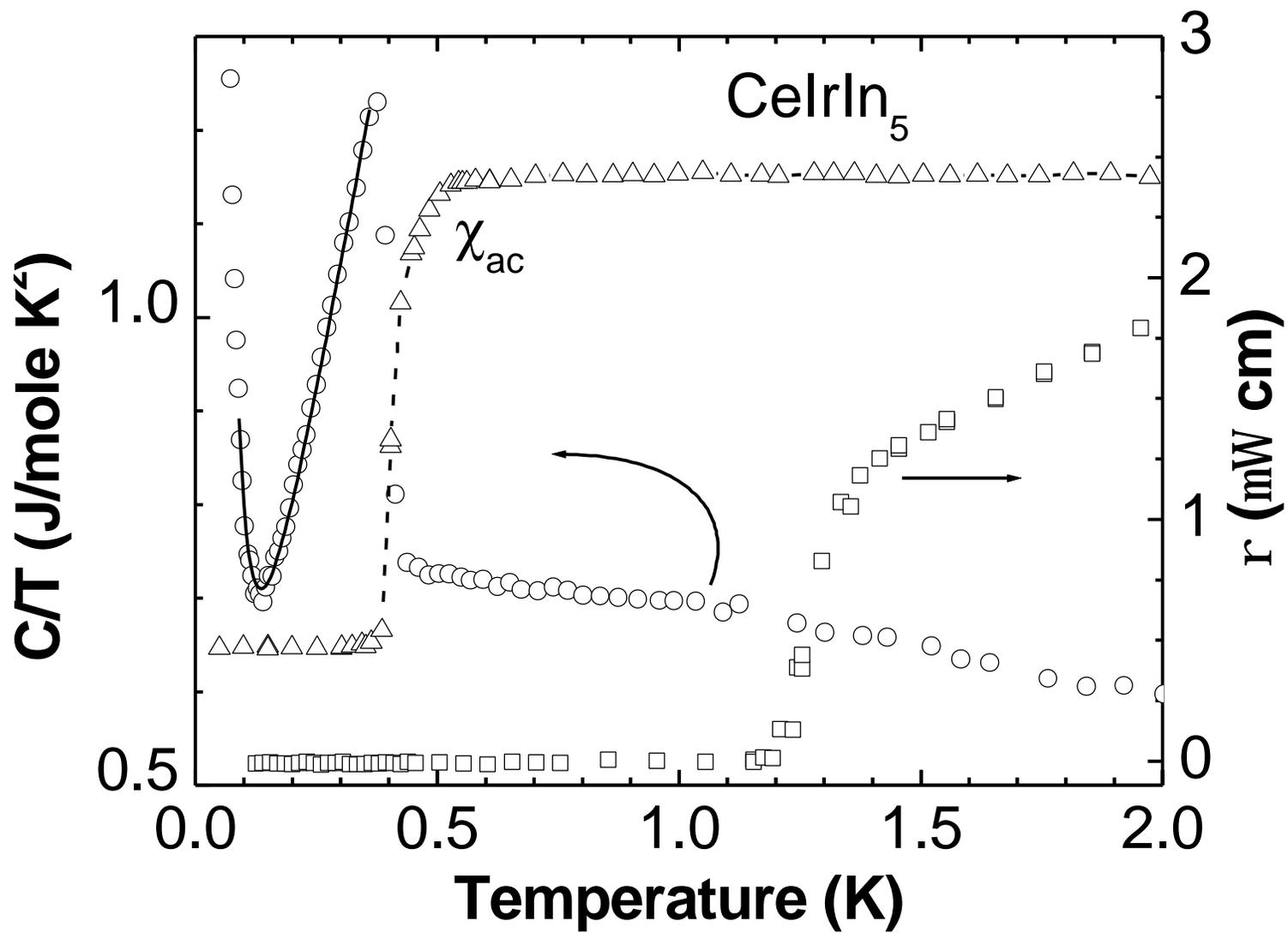

Figure 2 Petrovic, *et al*

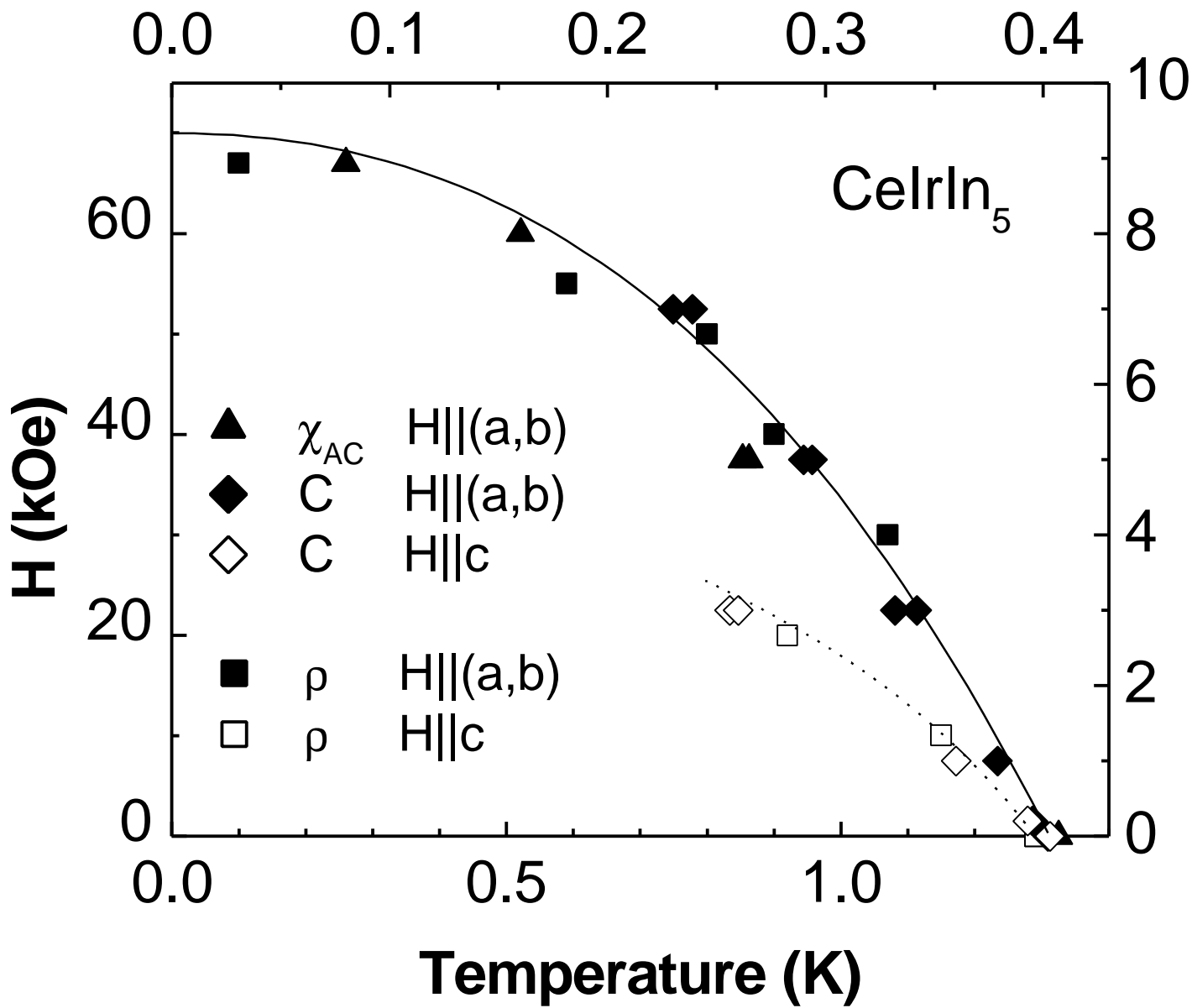

Figure 3 Petrovic, *et al*